\documentclass[twocolumn,showpacs,preprintnumbers]{revtex4}

\usepackage{amsmath}
\usepackage{graphicx}
\usepackage{dcolumn}
\usepackage{bm}

\begin{document}

\title{Solutions to the anisotropic quantum Rabi model}

 \author{Liwei Duan$^{1}$ and Qing-Hu Chen$^{1,2,*}$}

\address{
$^{1}$ Department of Physics, Zhejiang University, Hangzhou 310027,
P. R. China \\
$^{2}$  Collaborative Innovation Center of Advanced Microstructures, Nanjing University, Nanjing 210093, China
 }

 \date{\today }

\begin{abstract}
In this work, the anisotropic quantum Rabi model with different coupling
strengths of the rotating-wave and counter-rotating wave terms is studied
by using two kinds of extended coherent states (ECS). By the first kind of  ECS, we can derive a so-called $G$-function, by which both the regular and exceptional
solutions can be given. The exceptional solution are just corresponding to the  crossing points of two
energy levels with different parities, so is doubly degenerate.  By the second kind of  ECS, a general scheme for the
eigensolutions is derived analytically in a unified way. The zero-order
approximation is just the adiabatic approximation, and the first-order
approximation is actually a generalized rotating-wave approximation. The
algebraic formulae for the eigensolutions are given explicitly in  two
approximations. The generalized rotating-wave approximations work well in a
wide range of  two different coupling strengths and the qubit detunings.


\end{abstract}

\maketitle

\section{Introduction}

Quantum Rabi model (QRM) describes a two-level system coupled to a cavity
electromagnetic mode (an oscillator)~\cite{Rabi}, which can be used to
describe the simplest matter-light interactions, and has many applications
in numerous fields ranging from quantum optics, quantum information science
to condensed matter physics. In conventional quantum optics~\cite{Scully},
the rotating-wave (RW) terms are kept and the counter-rotating-wave (CRW)
terms are neglected. So usually the rotating-wave approximation (RWA) is
employed and analytical closed-form exact solutions are available.

Recently, in the circuit QED systems~\cite
{Wallraff,Deppe,Fink,Niemczyk,exp,Mooij}, the coupling between the
superconducting qubit and the resonator can be strengthened by $10\%$. In
this ultrastrong-coupling regime, the evidence for the breakdown of the RWA
has been provided by the transmission spectra~\cite{Niemczyk}. The
remarkable Bloch-Siegert shift associated with the counter-rotating terms
also demonstrates the failure of the RWA~\cite{exp}. So CRW terms could not
be omitted, and the full QRM should be considered. Although the numerical
solutions to the full QRM are extremely easy to obtain, the analytical
solutions are however highly nontrivial. The analytical approximate
solutions have been obtained at different levels, such as weak energy
difference $\left( \Delta \right) \;$between two-levels ($\Delta /\omega
<0.5,\omega \;$ is the cavity frequency) and deep strong coupling regime~
\cite{Casanova,Hausinger}, weak and intermediate coupling ($g/\omega <0.4$) ~%
\cite{He}, and whole parameter range~\cite{Feranchuk,Irish,zhengh,Beaudoin}.
The analytic exact solutions have been also obtained by many groups~\cite
{Swain,Chen2011, Braak, Chen2012}. Among these exact approaches, the
eigenvalues are usually (or equivalently) determined by zeros of the derived
functions. Two continued fraction techniques are formulated on the original
Fock space~\cite{Swain} and optimum extended coherent states (ECS)~\cite
{Chen2011}, where the built-in truncation is unavoidable formally. Recently,
Braak presented an analytical exact solution~\cite{Braak} using the Bargmann
representation. A so-called $G$-function has been derived, which actually
can be written in terms of Heun functions. Although it is not in a closed
form, a built-in truncation is not needed formally before the practical
calculation, in contrast with the continued fraction techniques.
Alternatively, using the method of extended coherent states, this solution
was recovered in a simpler and physically more intuitive way~\cite{Chen2012}.

The anisotropic matter-light interacting systems with different RW and CRW
coupling strengths have been studied for a long time, mostly for the
theoretical interest previously. The quantum chaos has been studied in the
anisotropic Dicke model~\cite{Furuya}. Recently, the well known Goldstone
and Higgs modes have been demonstrated in optical systems with only a few
(artificial) atoms inside a cavity, which can be described by a few qubit
QRM ~\cite{yejw2013}. More recently, the study of an anisotropic QRM~\cite
{Fanheng,Tomka} was motivated by the recent experimental progress. This
model can be mapped onto the model describing a two-dimensional electron gas
with Rashba ($\alpha _R,$ RW coupling relevant) and Dresselhaus ($\alpha _D,$
CRW coupling dependent) spin-orbit couplings subject to a perpendicular
magnetic field~\cite{Erlingsson}. These couplings can be tuned by an applied
electric and magnetic field, allowing the exploration of the whole parameter
space of the model. This model can directly emerge in both cavity QED~\cite
{Schiroa} and circuit QED~\cite{Wallraff}. For example in Ref.~\cite{Grimsmo}
a realization of the anisotropic QRM based on resonant Raman transitions in
an atom interacting with a high finesse optical cavity mode is proposed.

The exact solutions for the anisotropic QRM have been obtained using the
Bargmann representation~\cite{Fanheng,Tomka}. The $G$-function was obtained
by Xie \textsl{et al}.~\cite{Fanheng}, where both the regular and
exceptional eigenvalues can be obtained. The isolated exact solutions at the
level crossing was found by Tomka \textsl{et al}.~\cite{Tomka}. On the other
hand, the approximate analytic solutions with explicit expressions in a wide
parameter regime are not given in literature, to the best of our knowledge.

The paper is organized as follows. In Sec. II, we describe the model of the
anisotropic QRM. In Sec. III, by using ECS technique, we derive a new $G$%
-function to the anisotropic QRM resembling the compact one in the isotropic
model, giving not only the exact regular spectra, but also exceptional
solutions right at all the level crossing points. In Sec. IV, by another ECS
approach, we present a generalized rotating-wave approximation (GRWA) to the
anisotropic QRM, the formulae for the eigenenergies and the eigenstates are
explicitly given. A brief summary will be presented finally.

\section{Model}

The Hamiltonian of the anisotropic QRM can be described as follows~\cite
{Fanheng,Tomka}
\begin{equation}
H=\frac 12\Delta \sigma _z+a^{\dagger }a+g_1\left( a^{\dagger }\sigma
_{-}+a\sigma _{+}\right) +g_2\left( a^{\dagger }\sigma _{+}+a\sigma
_{-}\right) ,  \label{Hamiltonian}
\end{equation}
where $\Delta $ is qubit energy difference, $a^{\dagger }$ $\left( a\right) $
is the photonic creation (annihilation) operator of the single-mode cavity
with frequency $\omega $, $g_1\ $and $g_2\ $ are the RW and CRW coupling
constants respectively, and $\sigma _k(k=x,y,z)$ $\ $ are the Pauli
matrices. Set $r=g_2/g_1\ $as the anisotropic parameter.

\section{Analytical exact solutions within $G$-function technique}

Employing the following transformation
\begin{equation}
P=\frac 1{\sqrt{2}}\left(
\begin{array}{ll}
\sqrt{r} & ~1 \\
-\sqrt{r} & \;1
\end{array}
\right) ,\;\;P^{-1}=\frac 1{\sqrt{2}}\left(
\begin{array}{ll}
\;\frac 1{\sqrt{r}}\; & ~-\frac 1{\sqrt{r}} \\
\;\;\;\;1 & \;\;\;\;\;1
\end{array}
\right) ,  \label{P1}
\end{equation}
we have the Hamiltonian in the matrix (in units of $\hbar =\omega =1$)
\begin{widetext}
\begin{equation}
H_1=PHP^{-1}=\left(
\begin{array}{ll}
a^{\dagger }a+\beta \left( a+a^{\dagger }\right) +\left( \frac{\lambda _{+}}%
\beta -\beta \right) a^{\dagger } & \;\;\;\;-\frac 12\Delta -\frac{\lambda
_{-}}\beta a^{\dagger } \\
\;\;\;\;-\frac 12\Delta +\frac{\lambda _{-}}\beta a^{\dagger } &
\;a^{\dagger }a-\beta \left( a+a^{\dagger }\right) -\left( \frac{\lambda _{+}%
}\beta -\beta \right) a^{\dagger }
\end{array}
\right) ,
\end{equation}
\end{widetext}
where $\lambda _{\pm }=\left( g_1^2\pm g_2^2\right) /2$ and$\;\beta =\sqrt{%
g_1g_2}$.

We introduce two displaced bosonic operators with opposite displacements
\begin{equation}
A_{+}^{\dagger }=a^{\dagger }+\beta ;\;\;A_{-}^{\dagger }=a^{\dagger }-\beta
.  \label{dis}
\end{equation}
The bosonic number state in terms of the new photonic operators $%
A_{+}^{\dagger }$ and $A_{-}^{\dagger }$ are
\begin{eqnarray*}
\left| n\right\rangle _{A_{+}} &=&\frac{\left( A_{+}^{\dagger }\right) ^n}{%
\sqrt{n!}}D(-\beta )\left| 0\right\rangle \\
\left| n\right\rangle _{A_{-}} &=&\frac{\left( A_{-}^{\dagger }\right) ^n}{%
\sqrt{n!}}D(\beta )\left| 0\right\rangle ,
\end{eqnarray*}
where $D(\beta )=\exp \left( \beta a^{\dagger }-\beta a\right) $ is the
unitary displacement operator, $\left| 0\right\rangle $ is original vacuum
state, $\left| n\right\rangle _{A_{+}}$ and $\left| n\right\rangle _{A_{-}}$
are called ECS~\cite{chenqh}.

The Hamiltonian in terms of $A_{+}^{\dagger }$ can be written as
\begin{widetext}
\begin{equation}
H_1=\left(
\begin{array}{ll}
A_{+}^{\dagger }A_{+}+\left( \lambda _{+}/\beta -\beta \right)
A_{+}^{\dagger }-\lambda _{+} & \;\;\;\;\left( -\frac 12\Delta +\lambda
_{-}\right) -\frac{\lambda _{-}}\beta A_{+}^{\dagger } \\
\;\;\;\;\left( -\frac 12\Delta -\lambda _{-}\right) +\frac{\lambda _{-}}%
\beta A_{+}^{\dagger } & \;A_{+}^{\dagger }A_{+}-\left( \beta +\lambda
_{+}/\beta \right) A_{+}^{\dagger }-2\beta A_{+}+2\beta ^2+\lambda _{+}
\end{array}
\right) .  \label{H_G}
\end{equation}
\end{widetext}
The wavefunction can be expressed as the following series expansion using
the ECS
\begin{equation}
\left| A_{+}\right\rangle _{}=\left(
\begin{array}{c}
\sum_{n=0}^\infty \sqrt{n!}e_n|n\rangle _{A_{+}} \\
\sum_{n=0}^\infty \sqrt{n!}f_n|n\rangle _{A_{+}}
\end{array}
\right) .  \label{wave_A+}
\end{equation}
Projecting $\;_{A_{+}}\left\langle m\right| \;$ onto the Schr$\stackrel{..}{o%
}$dinger equation yields the recurrence relations for the coefficients
\begin{widetext}
\begin{eqnarray}
e_m &=&\frac{\left( \beta -\frac{\lambda _{+}}\beta \right) e_{m-1}+\left(
\frac 12\Delta -\lambda _{-}\right) f_m+\frac{\lambda _{-}}\beta f_{m-1}}{m-x%
},  \label{em} \\
f_m &=&\frac{\;\left( -\frac 12\Delta -\lambda _{-}\right) e_{m-1}+\frac{%
\lambda _{-}}\beta e_{m-2}+\left( m-1+2\beta ^2+2\lambda _{+}-x\right)
f_{m-1}-\left( \beta +\lambda _{+}/\beta \right) f_{m-2}}{2\beta m},
\label{fm}
\end{eqnarray}
\end{widetext}
where $x=\lambda _{+}+E$ ($E$ is the energy). Starting from $f_0=1,$ we can
obtain all $f_m\;$recursively, which will be very useful later.

Considering the conserved parity, by the coefficients in Eq. (\ref{wave_A+}%
), the wavefunction can also be expressed in the ECS of the $A_{-}$-space as
\begin{equation}
\left| A_{-}\right\rangle _{}=\left(
\begin{array}{c}
\sum_{n=0}^\infty \left( -1\right) ^n\sqrt{n!}f_n|n\rangle _{A-} \\
\sum_{n=0}^\infty \left( -1\right) ^n\sqrt{n!}e_n|n\rangle _{A-}
\end{array}
\right) .  \label{wave_A-}
\end{equation}
If both wavefunctions (\ref{wave_A+}) and (\ref{wave_A-}) are the true
eigenfunction for a non-degenerate eigenstate with eigenvalue $E$, they
should be in principle only different by a complex constant $r^{\prime }$
\begin{eqnarray}
\sum_{n=0}^\infty \sqrt{n!}e_n|n\rangle _A &=&r^{\prime }\sum_{n=0}^\infty
\left( -1\right) ^n\sqrt{n!}f_n|n\rangle _A;  \nonumber \\
\;\sum_{n=0}^\infty \sqrt{n!}f_n|n\rangle _A &=&r^{\prime }\sum_{n=0}^\infty
\left( -1\right) ^n\sqrt{n!}e_n|n\rangle _A.  \label{prop}
\end{eqnarray}
Left multiplying the original vacuum state ${\langle }0|$ to both side of
the above equations, and eliminating the ratio constant $r^{\prime }$ gives
\[
\sum_{n=0}^\infty e_n\beta ^n\sum_{n=0}^\infty e_n\beta ^n=\sum_{n=0}^\infty
f_n\beta ^n\sum_{n=0}^\infty f_n\beta ^n,
\]
where we have used
\begin{equation}
\sqrt{n!}{\langle }0|n{\rangle }_{A+}=(-1)^n\sqrt{n!}{\langle }0|n{\rangle }%
_{A-}=e^{-\beta ^2/2}\beta ^n,
\end{equation}
then we define the following $G$-function with the help of Eq. (\ref{em})
\begin{equation}
G_{\pm }(x)=\sum_{n=0}^\infty \left( f_n\mp e_n\right) \beta ^n,
\label{G-function}
\end{equation}
where $+(-)$ in the left-hand side is corresponding to even (odd) parity,
all coefficients are determined by Eqs. (\ref{em}) and (\ref{fm}). If $%
g_1=g_2=g$, the $G$-function of the isotropic QRM~\cite{Braak} is readily
recovered.

\begin{figure}[tbp]
\center
\includegraphics[width=7cm]{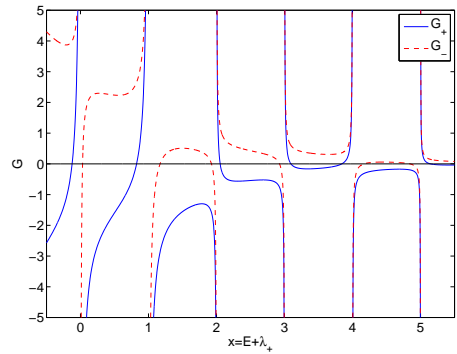}
\caption{ (Color online) G(x) curves for the anisotropic QRM at $%
\Delta=0.7,g_1=0.8,g_2=0.6$. The blue (red) curve denotes even (odd) parity.}
\label{G_function}
\end{figure}

\begin{figure}[tbp]
\center
\includegraphics[width=4cm]{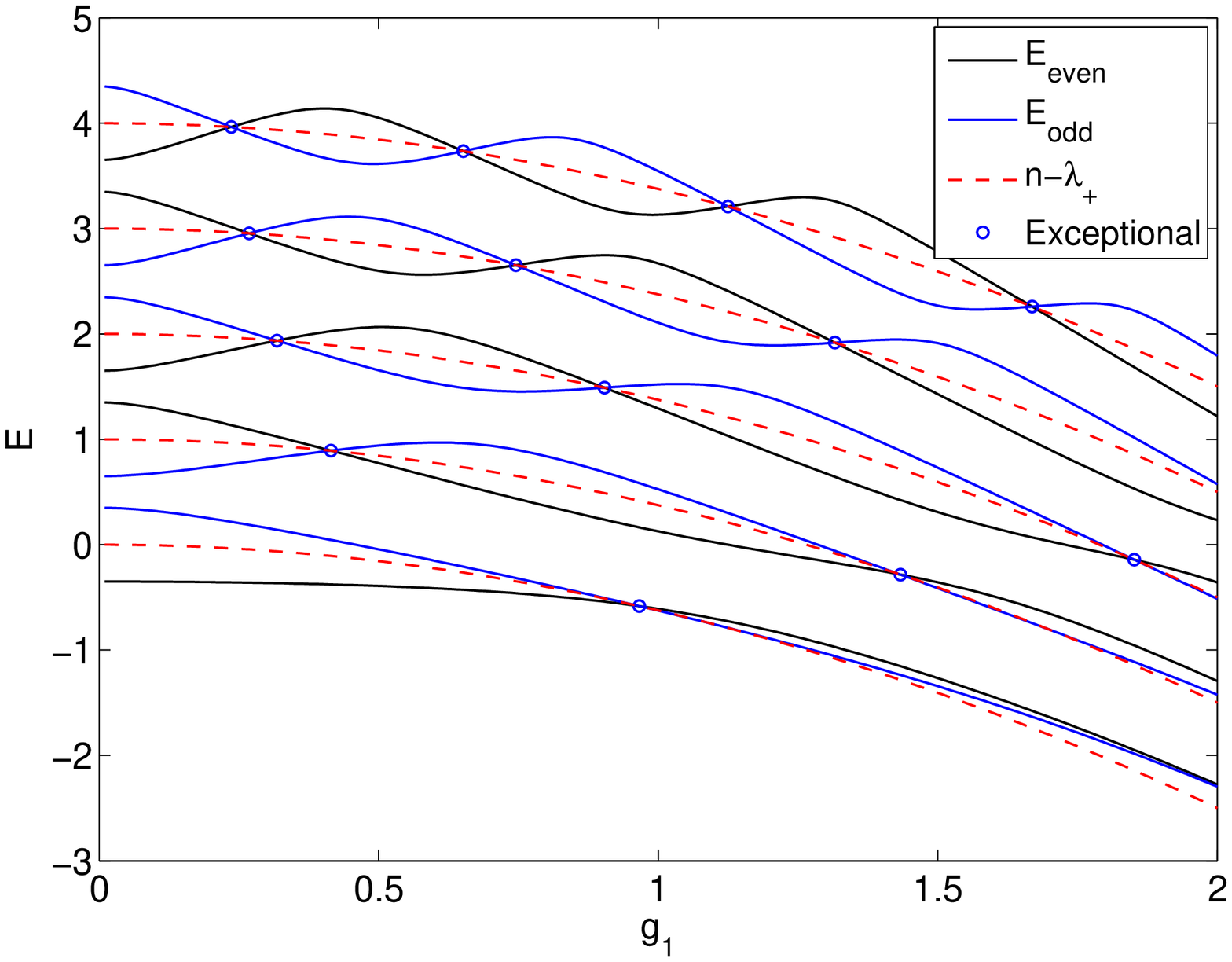} %
\includegraphics[width=4cm]{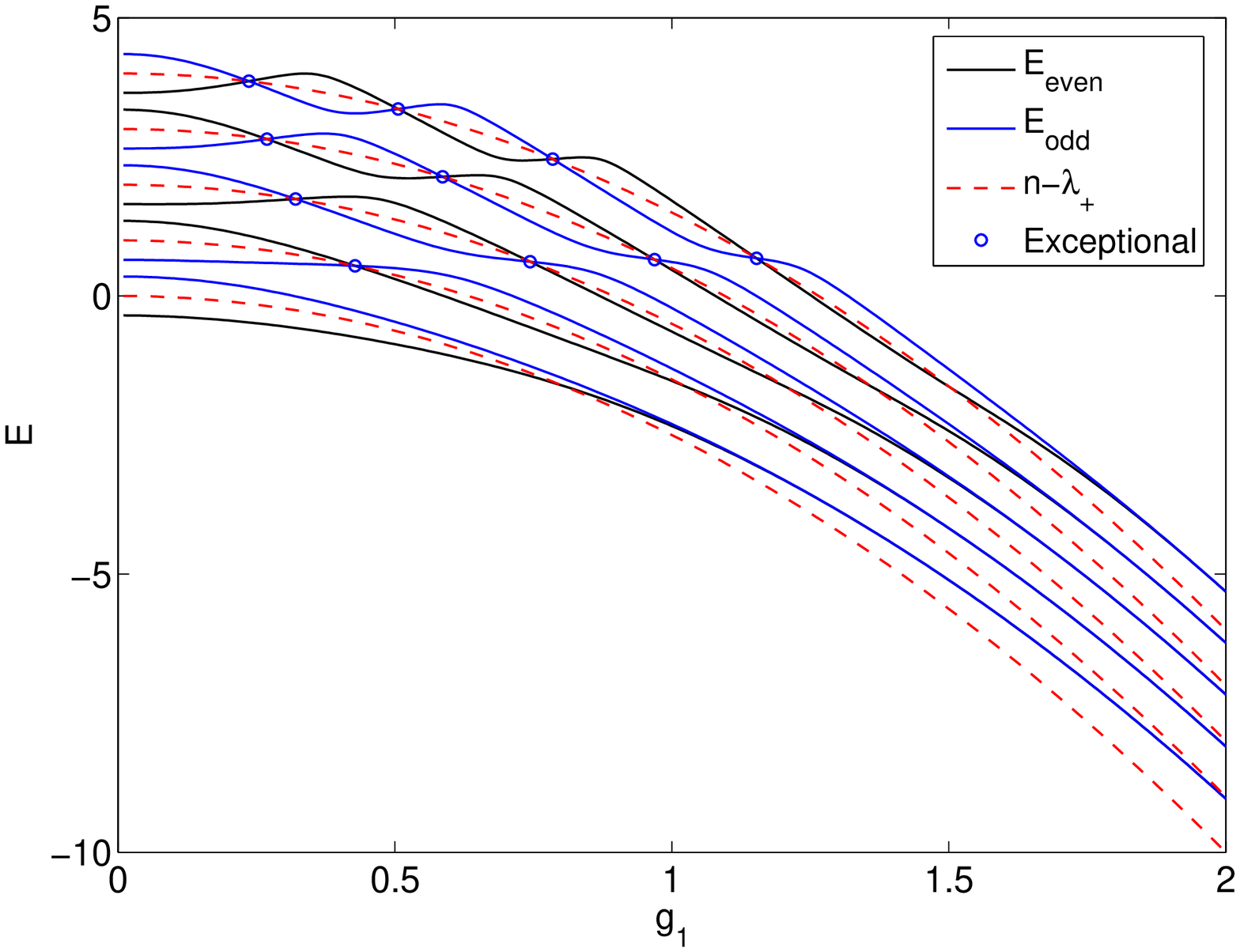}
\caption{ (Color online) The spectra and the isolated exceptional solutions
for the anisotropic QRM at $\Delta =0.7$. $r=1/2$ (left) $r=2$ (right). }
\label{energylevel}
\end{figure}

We plot the $G$-function for $\Delta =0.7,g_1=0.8,g_2=0.6$ in Fig.~\ref
{G_function}. The zeros reproduce all regular spectra, which can be
confirmed by the numerical exact solutions. The energy spectra for $\Delta
=0.7,r=1/2$ and $2$ are presented in Fig. \ref{energylevel}.

\textsl{Exceptional solutions:} We link the degenerate states to the Juddian
solutions~\cite{Judd}. Koc \textsl{et al}.~\cite{Koc} have obtained isolated
exact solutions in the isotropic QRM, which are just the Juddian solutions
with doubly degenerate eigenvalues. The degenerate eigenstates are excluded
in principle in the solutions based on the proportionality of Eq. (\ref{prop}%
) used in the present ECS technique. It naturally follows that the Juddian
solutions are exceptional ones. With the $G$-function (\ref{G-function}) at
hand, we can also discuss the Juddian solution~\cite{Judd} readily. The $G$%
-function is also not analytic in $x$ but has simple poles at $x=0,1,2...$
For special values of model parameters $g_1,g_2,\Delta $, there are
eigenvalues which do not correspond to zeros of Eq. (\ref{G-function});
these are the exceptional solutions. All exceptional eigenvalues are given
by the positions of the poles $x=n$
\begin{equation}
E=n-\lambda _{+}.  \label{ex_e}
\end{equation}
The necessary and sufficient condition for the occurrence of the eigenvalue
is immediately given by
\begin{equation}
\left( \beta -\frac{\lambda _{+}}\beta \right) e_{n-1}+\left( \frac 12\Delta
-\lambda _{-}\right) f_n+\frac{\lambda _{-}}\beta f_{n-1}=0,
\label{condition}
\end{equation}
which provides a constraint on the model parameters. They occur when the
pole of $G_{\pm }(x)$ at $x=n$ is lifted because its numerator in Eq. (\ref
{em}) vanishes. Note that this exceptional eigenvalue belongs to the states
with both even and odd parities, so it is doubly degenerate, and should be
at the level crossing points without exceptions.

For $n=0,x=0$

\[
\left( \frac \Delta 2-\lambda _{-}\right) f_0=0,
\]
so we at most have one exceptional eigenvalue for $n=0$ at
\begin{equation}
g_1^{(c)}=\sqrt{\frac \Delta {1-r^2}}.  \label{ex_gc}
\end{equation}
It follows that the first excited state and the ground state only intersects
for the CRW coupling weaker than RW coupling. The parity in the lowest
energy state will change in this case, so the first-order quantum phase
transitions occur at $g_1^{(c)}$, in sharp contrast with the isotropic QRM.
From Eq.(\ref{ex_gc}), the first level crossing in Fig. \ref{energylevel}
(left) should occur at $g_1^{(c)}=0.9661$, consistent with the numerical
calculations.

For $n=1,x=1$, the condition for the occurrence of the exceptional solutions
is
\begin{equation}
2\left( g_1^2+g_2^2\right) -1+\frac{\Delta ^2-\left( g_1^2-g_2^2\right) ^2}%
4+\frac 2{\frac \Delta {g_1^2-g_2^2}-1}=0.
\end{equation}
If $g_1=g_2=g,$ it reduces to
\begin{equation}
\frac 14\Delta ^2+\left( 4g^2-1\right) =0,
\end{equation}
which is exactly the same as that in the isotropic QRM.

For high order exceptional solutions, the condition for the occurrence of
the exceptional solutions is rather complicated, but can be numerically
estimated straightforwardly from Eq.(\ref{condition}) . All of them are at
the level crossing points with open circles in Fig. \ref{energylevel}.
Finally by the iteration in Eq. (\ref{em}) , we can formally write the
condition for the $n$-th exceptional solution in terms of all coefficient $%
f_{i\le n}$%
\begin{equation}
\sum_{i\le n}\Gamma _i\;f_i=0,
\end{equation}
where $\Gamma _i$ is only model parameters dependent.

\section{\textsl{Generalized Rotating-Wave Approximation}}

To facilitate the approximately analytical study, we use a unitary
transformation

\begin{equation}
U=\frac 1{\sqrt{2}}\left(
\begin{array}{ll}
1 & -~1 \\
1 & \;1
\end{array}
\right) ;\;\;U^{\dagger }=\frac 1{\sqrt{2}}\left(
\begin{array}{ll}
\;1\; & ~1 \\
-1 & \;1
\end{array}
\right) ,  \label{U2}
\end{equation}
the Hamiltonian (\ref{Hamiltonian}) becomes
\begin{eqnarray}
H_2 &=&U^{\dagger }HU  \nonumber \\
&=&\left(
\begin{array}{ll}
a^{\dagger }a+\alpha \left( a^{\dagger }+a\right) & -\frac \Delta 2-\gamma
\left( a^{\dagger }-a\right) \\
-\frac \Delta 2+\gamma \left( a^{\dagger }-a\right) & \;a^{\dagger }a-\alpha
\left( a^{\dagger }+a\right)
\end{array}
\right)  \label{H_2}
\end{eqnarray}
where$\;\alpha =\left( g_1+g_2\right) /2;\;\gamma =\left( g_1-g_2\right) /2$.

Two displaced bosonic operators with opposite displacements, different from
those in Sec. III, are introduced
\begin{equation}
B_{+}^{\dagger }=a^{\dagger }+\alpha ,\;B_{-}^{\dagger }=a^{\dagger }-\alpha
.  \nonumber
\end{equation}
The number state for the bosonic particles $B_{+}$ and $B_{-}$ are
\begin{equation}
\left| n\right\rangle _{B_{+}}=\frac{\left( B_{+}^{\dagger }\right) ^n}{%
\sqrt{n!}}D(-\alpha )\left| 0\right\rangle ,\;\left| n\right\rangle _{B_{-}}=%
\frac{\left( B_{-}^{\dagger }\right) ^n}{\sqrt{n!}}D(\alpha )\left|
0\right\rangle ,
\end{equation}
then the Hamiltonian can be expressed with the number operators of the
particles $B_{+}$ and $B_{-}$%
\begin{equation}
H_2=\left(
\begin{array}{ll}
\;\;\;B_{+}^{\dagger }B_{+}-\alpha ^2 & \;-\frac \Delta 2-\gamma \left(
B_{-}^{\dagger }-B_{-}\right) \\
-\frac \Delta 2+\gamma \left( B_{+}^{\dagger }-B_{+}\right) &
\;\;\;\;\;\;\;B_{-}^{\dagger }B_{-}-\alpha ^2
\end{array}
\right) .
\end{equation}
The wavefunction can be expressed as the following series expansion using
these ECS
\begin{equation}
\left| B\right\rangle =\left(
\begin{array}{l}
\;\;\;\;\sum_{n=0}^{N_{tr}}\sqrt{n!}c_n\left| n\right\rangle _{B_{+}} \\
\pm \sum_{n=0}^{N_{tr}}\sqrt{n!}(-1)^nc_n\left| n\right\rangle _{B_{-}}
\end{array}
\right) ,  \label{wavefunction_p}
\end{equation}
where $+\left( -\right) $ stands for even (odd) parity, and $N_{tr}$ is the
truncated number of particles $B_{+}$ and $B_{-}$.

Projecting  $|m\rangle _{B+}\;$onto the Schr$\stackrel{..}{o}$dinger
equation gives
\begin{equation}
\left( m-\alpha ^2-E\right) c_m\mp (-1)^m\sum_{n=0}^{N_{tr}}\;R_{m,n}c_n=0,
\label{matrix}
\end{equation}
where
\begin{equation}
R_{m,n}=\frac \Delta 2\;D_{m,n}-\gamma \left( D_{m,n+1}-nD_{m,n-1}\right) ,
\label{Rmn}
\end{equation}
where
\begin{eqnarray}
&& D_{mn} = _{B_{+}}\left\langle m\right| \sqrt{\frac{n!}{m!}}%
(-1)^{n-m}\left| n\right\rangle _{B_{-}}  \nonumber \\
&& = (-1)^m\exp (-2\alpha ^2)\sum_{k=0}^{\min [m,n]}(-1)^k\frac{n!(2\alpha
)^{m+n-2k}}{(m-k)!(n-k)!k!} \nonumber
\end{eqnarray}
Note that $\;D_{mn}=\;\left( 2\alpha \right) ^{n-m}\exp (-2\alpha
^2)L_m^{n-m}(4\alpha ^2)$ for $m\le n$,\ $D_{mn}=\frac{n!}{m!}\;(-2\alpha
)^{m-n}\exp (-2\alpha ^2)L_n^{m-n}(4\alpha ^2)\;$for $\;m\ge n$, and $%
D_{m,n}=0$ if $m<0$ or $n<0$. Here $L(y)$ is the Laguerre polynomial. Eq. (%
\ref{matrix}) can be reduced to that in the isotropic QRM \cite{Qinghu} if
set $\gamma =0,r=1\;(i.e.\;\alpha =g)$. Next, we will perform the
approximation step by step.

\textsl{Adiabatic approximations}: if $N_{tr}=0$, i.e. the zero-order
approximation, the eigenfunctions for the quantum number $m\;$are give by ($%
k=1,2$)
\begin{equation}
\left| km\right\rangle ^{(0)}=\left(
\begin{array}{l}
\left| m\right\rangle _{B+} \\
\pm \left( -1\right) ^m\left| m\right\rangle _{B-}
\end{array}
\right) ,
\end{equation}
and the corresponding eigenvalues are
\begin{equation}
E_m^{(k)}=m-\alpha ^2\mp (-1)^mR_{m,m},  \label{ZOA}
\end{equation}
Similar to the isotropic QRM~\cite{Qinghu}, the zero-order approximation in
this technique is just the adiabatic approximations. It is also the same as
the adiabatic approximation in Ref. \cite{Tomka} derived in an alternative
way. Note that in the adiabatic approximations, the transition between
states belonging to different manifolds $m$ is neglected.

\textsl{Generalized Rotating-Wave Approximations}: Beyond the adiabatic
approximation, the transition between different manifolds should be
considered. We will perform a further correction by taking into account the
transition between states belonging to two manifolds $m$ and $m+1$. The
solutions for main quantum number $m\left( =0,1,2..\right) \ $can be
obtained by selecting two terms in Eq. (\ref{matrix}) for each $m$ and $m+1$.
Considering $\mp \left( -1\right) ^m=1$ for the implied parity,
we have  the following determinant in a $2$-by-$2$ block  starting with $m$
\[
\left|
\begin{array}{ll}
\;\left( -x-\alpha ^2+R_{m,m}\right)  & \;\;\;\;\;\;\;\;\;R_{m,m+1} \\
\;\;\;\;\;\;\;\;\;-R_{m+1,m} & \left( 1-x-\alpha ^2-R_{m+1,m+1}\right)
\end{array}
\right| =0,
\]
where $E=m+x$. So we readily obtain the eigenenergies for each $m\;\left( =0,1,2...\right) $
as
\begin{widetext}
\begin{equation}
E_m^{(k)}=m+\frac 12-\alpha ^2+\frac 12\left( R_{m,m}-R_{m+1,m+1}\right)
+\left( -1\right) ^k\frac 12\sqrt{\left( \left[ 1-\left(
R_{m,m}+R_{m+1,m+1}\right) \right] ^2-4R_{m,m+1}R_{m+1,m}\right) },
\label{energy_GRWA}
\end{equation}
\end{widetext}
Note that for isotropic QRM, it is exactly the same as the GRWA result
derived in Ref. \cite{Irish,Feranchuk} in the isotropic QRM. So the present
first-order approximation is also termed as the GRWA.

The lowest energy with even parity is
\begin{eqnarray}
E_0^{(even)} &=&\frac 12-\alpha ^2+\frac 12\left( R_{0,0}-R_{1,1}\right)
\nonumber \\
&&\pm \frac 12\sqrt{\left( \left[ 1-\left( R_{0,0}+R_{1,1}\right) \right]
^2-4R_{0,1}R_{1,0}\right) }.  \label{E_0}
\end{eqnarray}

\begin{figure}[tbp]
\includegraphics[width=4cm]{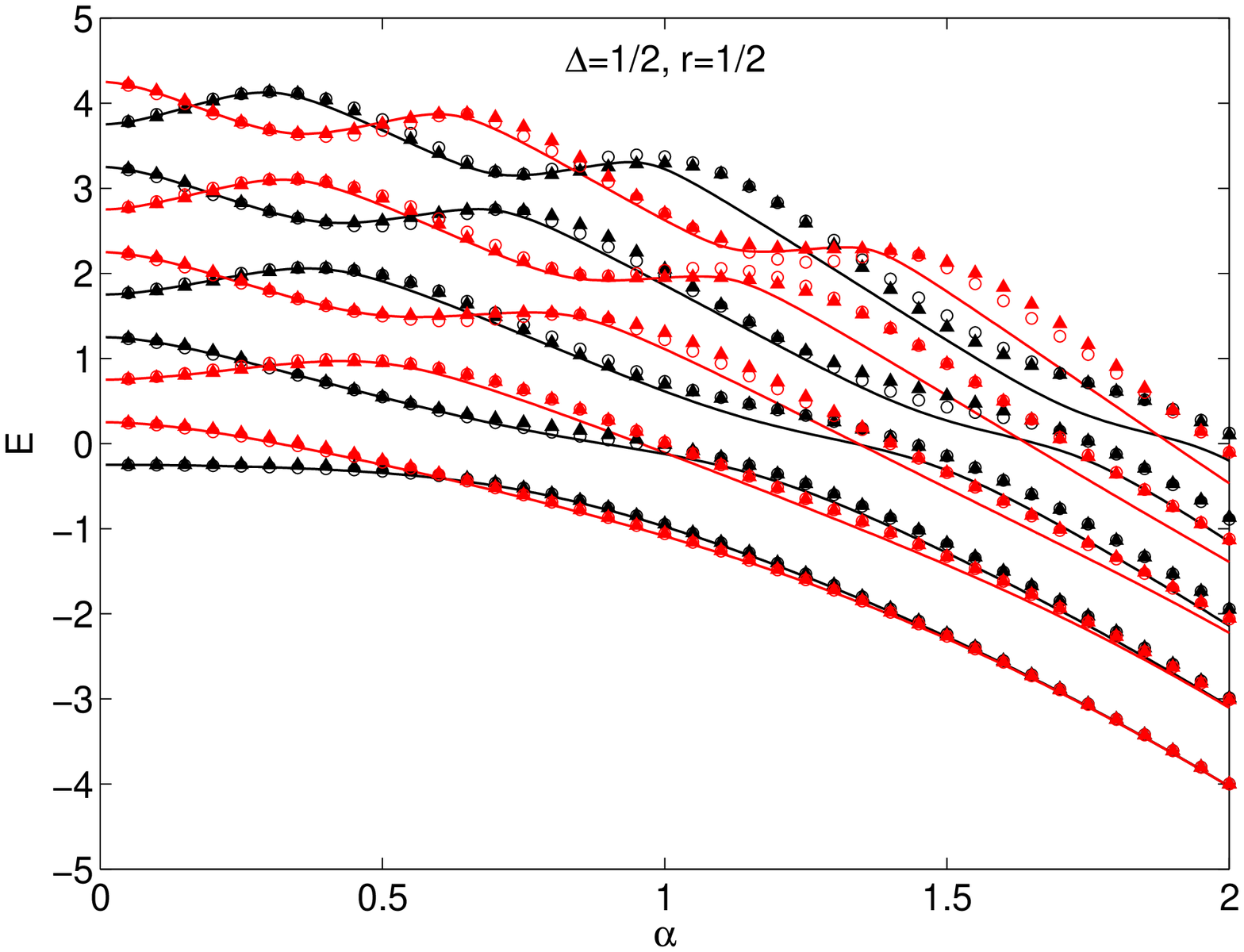} %
\includegraphics[width=4cm]{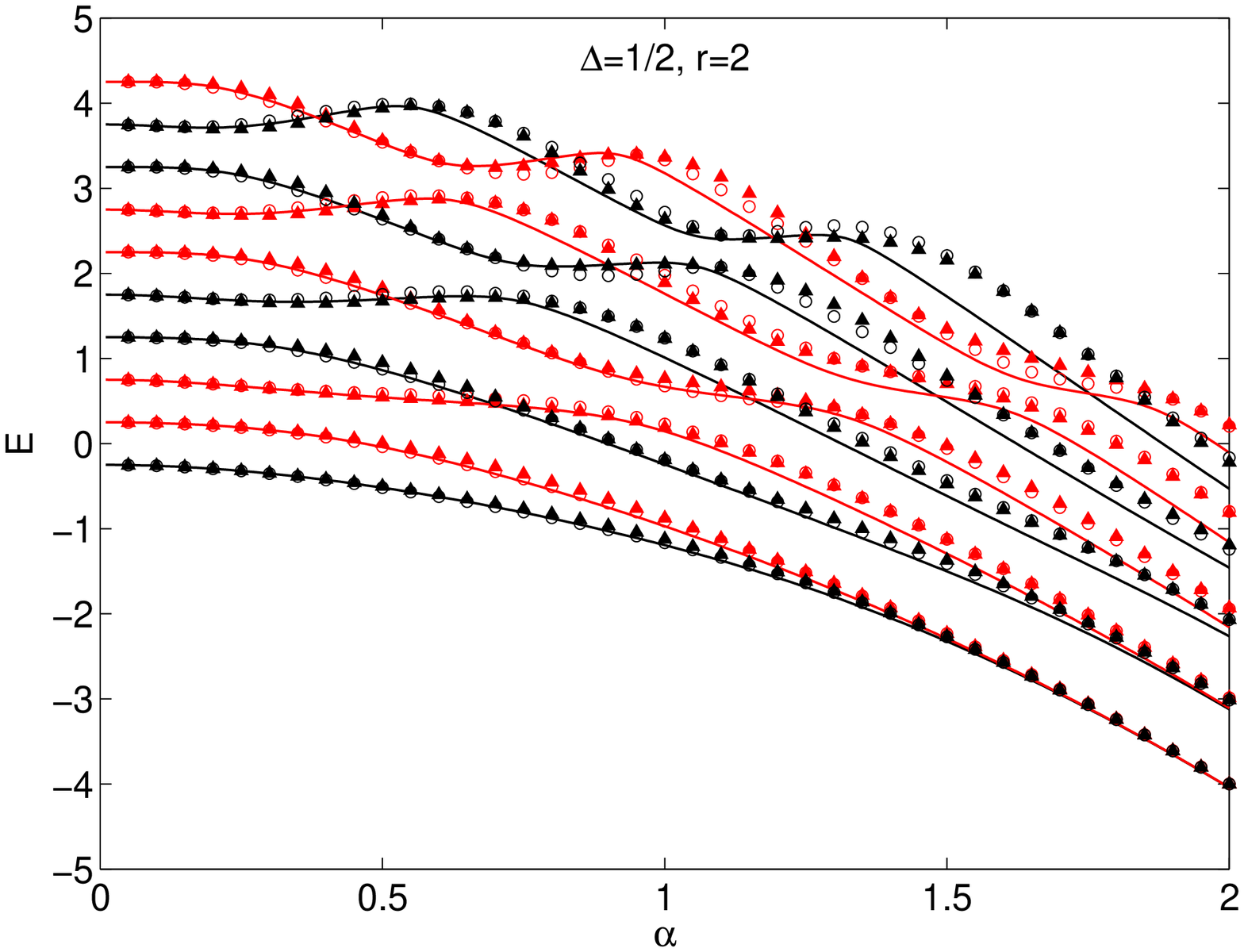}
 \includegraphics[width=4cm]{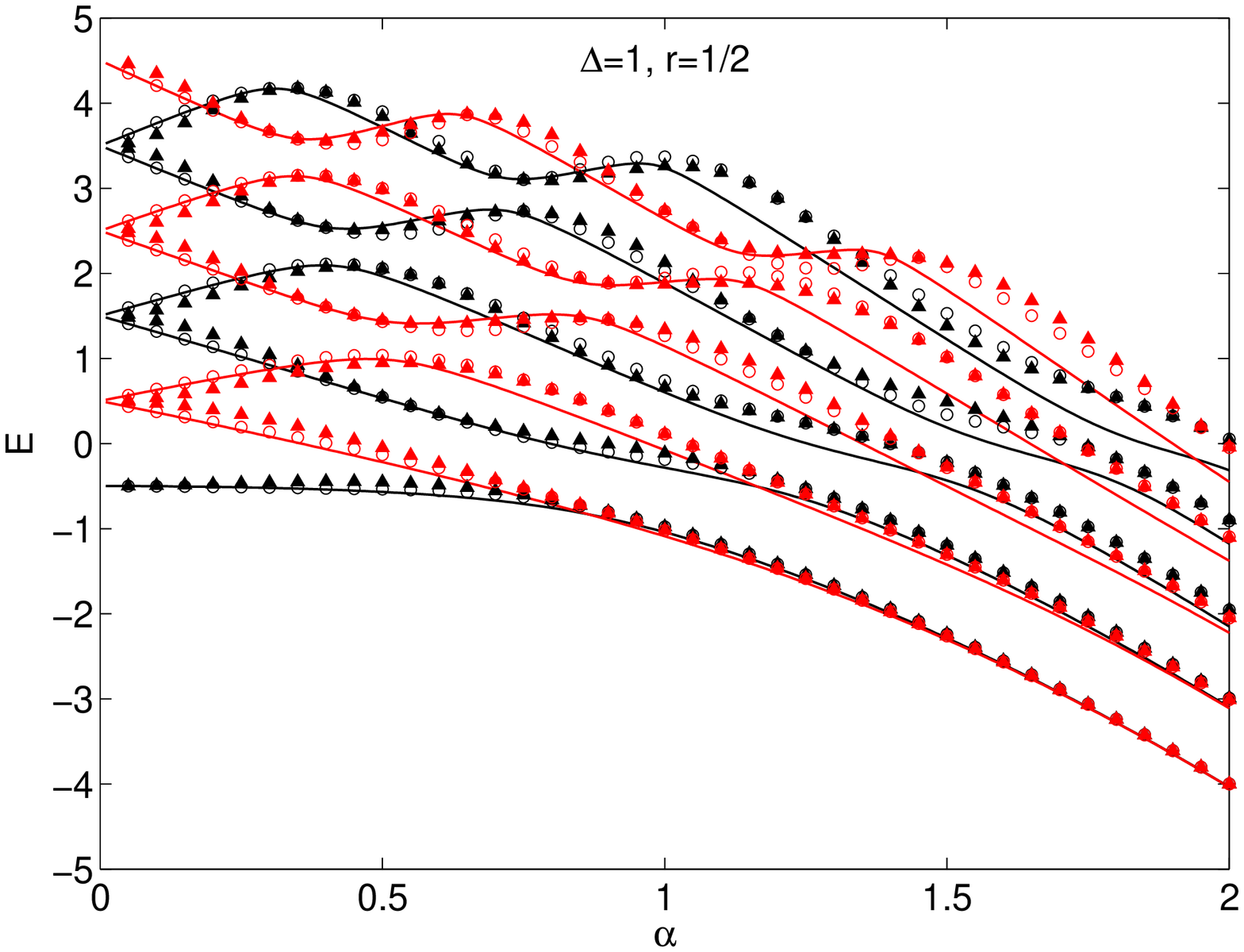}
\includegraphics[width=4cm]{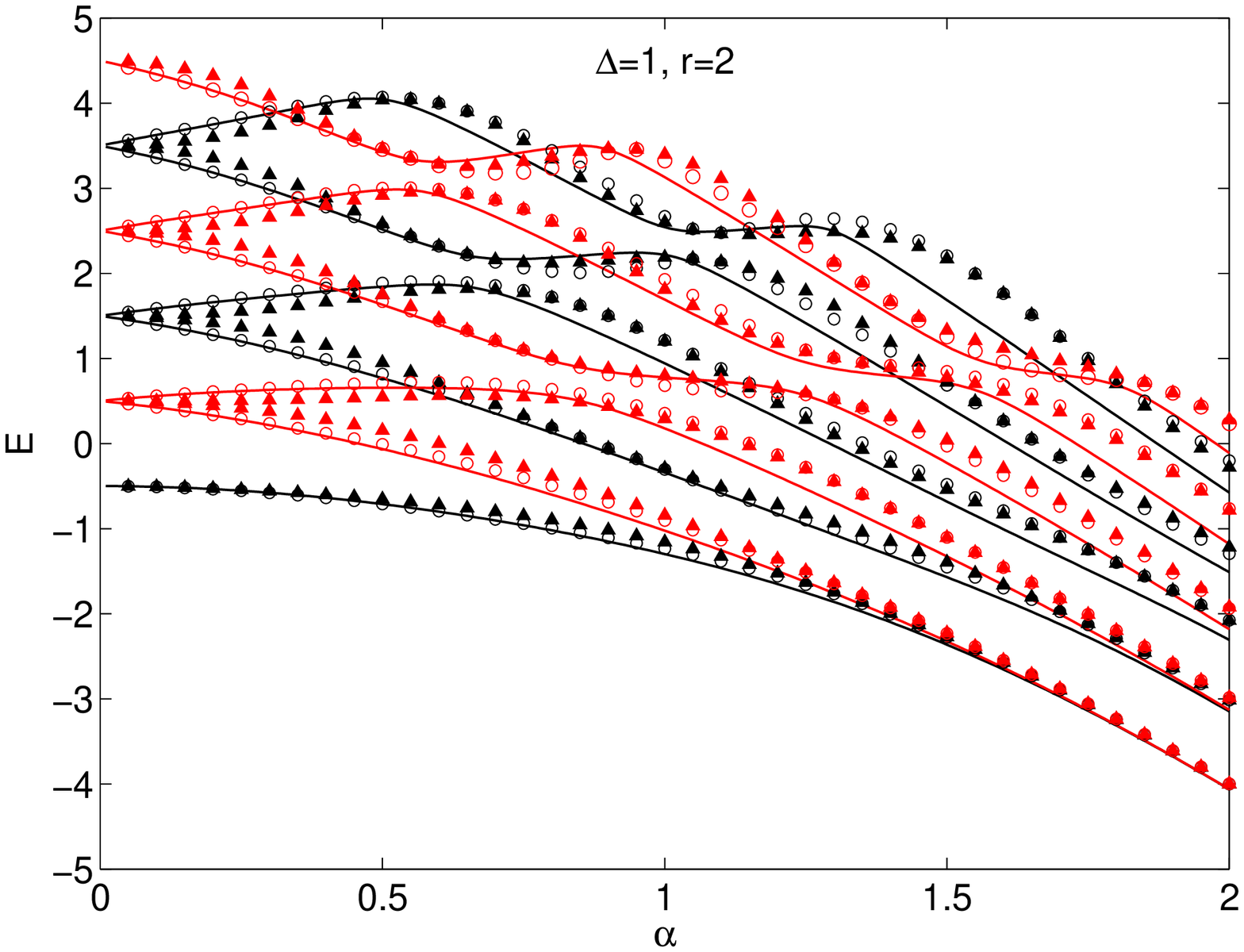}
\caption{ (Color online) The energy levels as a function of coupling
constant $\alpha =\left( g_1+g_2\right) /2$ for different anisotropic
parameters $r=1/2$ (left column) and $r=2$ (right column) and different
qubit splitting $\Delta =0.5$ (upper panel) and $1.0$ (down panel). GRWA
results are denoted by open circles, adiabatic ones by solid triangles,
exact ones by solid lines. Results for even and odd parities are
distinguished by black and red lines. }
\label{compare_r}
\end{figure}

For a given anisotropic parameters $r=1/2$ and $r=2$, by Eqs. (\ref
{energy_GRWA}) and Eq. (\ref{E_0}), we calculate the energy levels against $%
\alpha =\left( g_1+g_2\right) /2$ both in the GRWA and the adiabatic
approximation, which are presented in Fig. \ref{compare_r}. The exact ones
from the Sec. III are also presented for comparison. Obviously, the GRWA
results agree well with the exact ones qualitatively in the whole coupling
regime. The crossing properties are all present in the GRWA. The results by
the adiabatic approximation obviously deviate from the exact ones and become
worse with increasing $\Delta $ and decreasing $\alpha $. The reason is that
the transitions between states belonging to different manifolds in the true
physical process are neglected in the adiabatic approximation, but the
dominate transition from the manifolds $m$ and $m+1$ are taken into account
in the GRWA. From the energy spectra, the difference between these two
approximations is not very large, but they are essentially different. If the
wavefunction is involved in some physical process, the difference should be
remarkable. Further corrections to the GRWA will only result in qualitatively
different results, and will not considered here.

Due to the counter-rotating wave terms, the eigenfunctions and eigenvalues
of the anisotropic QRM present an open problem because they are not known in
anything like a closed form, even the exact solutions reported recently\cite
{Fanheng} and the present new $G$-functions. No analytical explicit
expressions for the exact eigenvalues for the whole coupling range are
available in the literature, to the best of our knowledge. The analytical
explicit expressions presented in this paper might be practically useful.

\section{Conclusions}

In this work, we first derive a concise $G$-function, resembling to the
compact one in the isotropic model, for the anisotropic QRM by using ECS,
then obtain quite accurate approximate analytical solutions by another ECS.
Zeros of the $G$-function will yield the regular spectra. The isolated exact
solutions are given by the exceptional solutions to this $G$-function. The
condition for their occurrence are also derived in the closed form. The
crossing points of the energy levels satisfy this condition, similar to the
single-mode QRM. The present analytic solution is well defined
mathematically, because of no built-in truncations, thereby allowing a
conceptually clear, practically feasible treatment to energy spectra and
many physical processes. The explicit expressions for the eigensolutions in
the GRWA are also obtained analytically by the another ECS. In a wide
coupling regime, the GRWA results are very close to the exact ones.

Interestingly, this work adds the anisotropic QRM to a list with a compact $%
G $-function like
\begin{equation}
G_{\pm }\left( x\right) =\sum_{n=0}^\infty f_n\left( 1\pm \frac{\sum_{i\le
n}\Gamma _if_i/f_n}{n-x}\right) L_n(g),  \label{universal}
\end{equation}
where $f_n^{}$ is determined recursively from $f_0=1$. For the isotropic QRM
with one-photon~\cite{Braak}, $\Gamma _i=\frac \Delta 2\delta
_{n,i},\;L_n=g^n$. For the isotropic QRM with two-photons~\cite{Chen2012}, $%
\Gamma _i=\frac \Delta {2\sqrt{1-4g^2}}\delta _{n,i}$, $L_n(g)$ is given by
Eqs. (48) and (49) there. For two-mode QRM~\cite{chenqh2015}, $\Gamma
_i=\frac \Delta {4\sqrt{1-g^2}}\delta _{n,i}$ and $L_n(g)$ is given by Eq.
(28) there. In the present model, $\Gamma _i$ is dependent on model
parameters and presents for all $i\le n,$ and $L_n=\left( \sqrt{g_1g_2}%
\right) ^n$. The qubit-cavity model possessing a compact $G$-function like
Eq. (\ref{universal}) shares the common property. The denominator of the
parity dependent term, i.e. the second term in Eq. (\ref{universal}), is $%
n-x $, so the zeros of its numerator will yield the condition for the
occurrence of the exceptional solutions: isolated doubly degenerate
eigenstates with eigenenergy $x(E)=n$, which is very helpful to analyze the
structure of the energy spectra. This list may be expanded by absorbing
other related models in the future.

\textbf{ACKNOWLEDGEMENTS}: This work was supported by National Natural Science Foundation of China
under Grant No. 11174254.

$^{*}$ Corresponding author. Email:qhchen@zju.edu.cn

\end{document}